\documentclass[preprint]{vgtc}               

\graphicspath{{figures/}{pictures/}{images/}{./}} 

\usepackage{times}                     

\usepackage{tabu}                      
\usepackage{booktabs}                  
\usepackage{lipsum}                    
\usepackage{mwe}                       

\usepackage{mathptmx}                  
\usepackage{amsmath}

\onlineid{1108}

\vgtccategory{Research}

\vgtcinsertpkg

\usepackage{pgfplots}
\usetikzlibrary{calc}
\pgfplotsset{compat=newest}

\usepackage{cool}
\usepackage{xcolor-solarized}

\newcommand{\mySetNotation}[1]{{\mathbb{#1}}}
\newcommand{\RRSet}{{\mySetNotation{R}}}

\newcommand{\myBoldNotation}[1]{{\mathbf #1}}

\newcommand{\ve}   {{\myBoldNotation{e}}}

\newcommand{\vn}   {{\myBoldNotation{n}}}

\newcommand{\vp}   {{\myBoldNotation{p}}}

\newcommand{\vx}   {{\myBoldNotation{x}}}

\newcommand{\mA}   {{\myBoldNotation{A}}}

\newcommand{\mI}   {{\myBoldNotation{I}}}

\newcommand{\mT}   {{\myBoldNotation{T}}}

\Style{IdentityMatrixSymb=\mI,%
       DSymb={\mathrm d},%
       IntegrateDifferentialDSymb={\mathrm d},%
       TrParen=p,%
       DetParen=p}
\Style{DDisplayFunc=outset}
\renewcommand{\Transpose}[1]{ {#1}^\Transp }

\newcommand{\Transp}{{{\mathrm T}}}

\title{Uncertain Mode Surfaces in 3D
Symmetric Second-Order Tensor Field Ensembles}

\author{Tim Gerrits\thanks{e-mail: gerrits@vis.rwth-aachen.de}}
\affiliation{\scriptsize RWTH Aachen University, Germany}

\teaser{
  \centering
    \begin{tikzpicture}
    \tikzstyle{image} = [inner sep=0, outer sep=0, node distance=0 and 0]
    \node[image] (image) at (0,0) {
        \includegraphics[width=\linewidth]{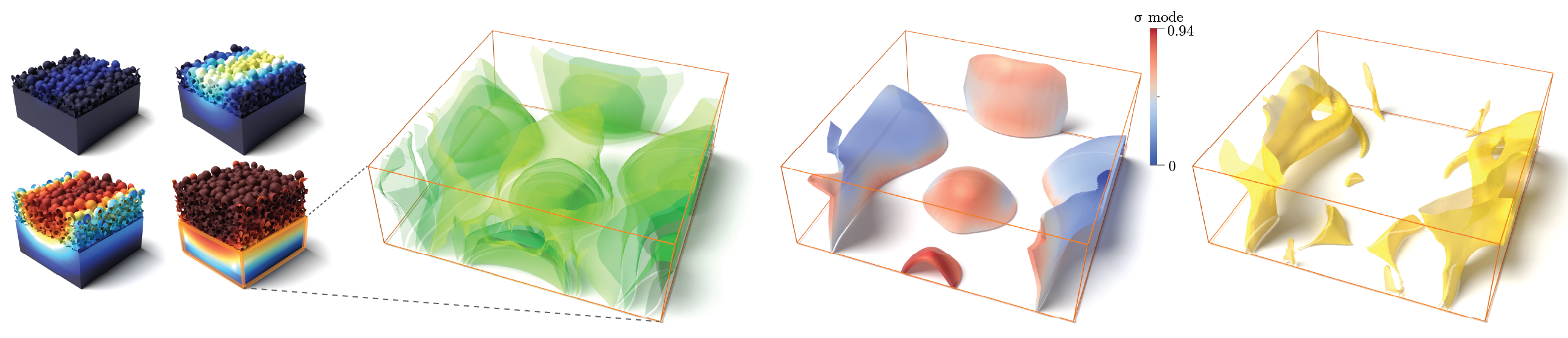}%
    };
    \begin{scope}[
        shift=(image.south west), 
	x={($(image.south east)-(image.south west)$)}, 
	y={($(image.north west)-(image.south west)$)}]
      \node[font=\small, text=black] at (0.1, 0.03) {a)};
      \node[font=\small, text=black] at (0.33, 0.03) {b)};
      \node[font=\small, text=black] at (0.6, 0.03) {c)};
      \node[font=\small, text=black] at (0.87, 0.03) {d)};
    \end{scope}
  \end{tikzpicture}
  \caption{Uncertain mode surface features in an elasto-plastic residual stress tensor ensemble from a multilayer physics simulation \cite{yang2024elasto} for the analysis of additive manufactured materials.
    a) The four ensemble members vary in powder scale, which influences the morphology of the powder bed (lower part).
    b) $-0.3$-mode surfaces extracted for each ensemble member represented by a different shade of green.
    c) The enhanced ensemble \textit{meanSurface} $\mathcal{M}_{-0.3}$ summarizing the $-0.3$-mode surfaces from all members, while its color and thickness encode the mode standard deviation, thus indicating the uncertainty of the feature.
    d) The generalized \textit{probabilityBand} indicates locations with a $33\%$ chance of mode values among the ensemble falling within the range of $-0.1$ to $-0.5$.
  }
  \label{fig:teaser}
}

\abstract{%
    The analysis of 3D symmetric second-order tensor fields often relies on topological features such as degenerate tensor lines, neutral surfaces, and their generalization to mode surfaces, which reveal important structural insights into the data.
    However, uncertainty in such fields is typically visualized using derived scalar attributes or tensor glyph representations, which often fail to capture the global behavior.
    Recent advances have introduced uncertain topological features for tensor field ensembles by focusing on degenerate tensor locations.
    Yet, mode surfaces, including neutral surfaces and arbitrary mode surfaces are essential to a comprehensive understanding of tensor field topology.
    In this work, we present a generalization of uncertain degenerate tensor features to uncertain mode surfaces of arbitrary mode values, encompassing uncertain degenerate tensor lines as a special case.
    Our approach supports both surface and line geometries, forming a unified framework for analyzing uncertain mode-based topological features in tensor field ensembles.
    We demonstrate the effectiveness of our method on several real-world simulation datasets from engineering and materials science.
}


\begin{document}


\firstsection{Introduction}

\maketitle

Various scientific fields rely on the analysis of symmetric 3D second-order tensor fields, including material science and medical imaging.
Topological features, such as degenerate tensor lines~\cite{hesselink97topo} and neutral surfaces~\cite{palacios2015feature} enable analysis and interpretation of global behavior of single -- ``certain" -- tensor fields.
However, uncertainty is an inherent aspect of many scientific pipelines~\cite{kamal2021recent} including simulations and measurements.
This is commonly captured by generating tensor field ensembles, typically collections of simulation runs with varying parameters.
Despite their importance, only a few techniques exist to visualize and analyze the topological behavior of entire tensor field ensembles.
Recent work by Schmitz and Gerrits~\cite{schmitz2024exploring} introduced a number of features indicating the location and likelihood of degenerate tensors within the ensemble based on different descriptions of the tensor mode -- a tensor invariant -- of both the mean tensor field and the distribution of mode values across the ensemble members.
Their approach, however, made limiting assumptions such as ignoring the mode sign and exclusively targeting degenerate tensor locations, thus offering no insight into the distribution of general mode surfaces within the ensemble.
As demonstrated by Qu et al.~\cite{qu2021mode}, the analysis of such surfaces can reveal important patterns and transitions in underlying physical processes captured by the tensor fields.

In this work, we present a generalization of uncertain degenerate tensor features to uncertain mode surfaces of arbitrary mode values.
We introduce three visualization techniques, namely the (enhanced) ensemble \textit{meanSurface}, the ensemble \textit{modeShell}, and the generalized ensemble \textit{probabilityBand}.
Similar to the original work, these features complement each other in communicating likely locations of mode surfaces along with a description of their uncertainty.
By explicitly incorporating mode sign, our method enables more nuanced analysis of tensor field ensembles while reducing to familiar line-based representations in cases of degenerate tensor locations.
We demonstrate this on two real-world simulation datasets from mechanical engineering and material science.
To the best of our knowledge, this presents the first unified approach for uncertainty-aware visualization of mode-based tensor features, combining both degenerate tensor locations and general mode surfaces.

\section{Background}
A 3D second-order tensor $\mT$ in an orthonormal basis can be represented by a $3 \times 3$ matrix and decomposed into a set of eigenvectors $\ve_i$ with $i \in [1,2,3]$ and corresponding eigenvalues $\lambda_1 \geq \lambda_2 \geq \lambda_3$, commonly referred to as major, medium, and minor eigenvalues.
A tensor is considered \textit{neutral} if the average of the major and minor eigenvalues equals the medium one, and \textit{degenerate} if at least two eigenvalues are equal. 
The tensor is referred to as being \textit{linear degenerate} if the major eigenvalue is unique, and \textit{planar degenerate} if the minor eigenvalue is. 
A tensor is called \textit{symmetric} if $\mT = \Transpose{\mT}$ and \textit{traceless} if the sum of its diagonal elements is zero, i.e., $tr(\mT) = 0$.
Throughout this work, we refer to \textit{symmetric second-order 3D tensors} simply as \textit{tensors}.
The determinant $\operatorname{det}(\mT)$ is given by the product of its eigenvalues, and any tensor can be decomposed into a traceless deviatoric tensor $\mA$ and a scalar multiple of the identity $\mI$ by $\mA = \mT - \frac{tr(\mT)}{3}\mI$.
The \textit{mode} of a tensor can be defined as
\begin{equation}
    \mu(\mT) = 3\sqrt{6}\operatorname{det}\frac{\operatorname{det}(\mA)}{||\mA||^3}
\end{equation}
where $\mu \in [-1,1]$.
Mode values are tightly linked to eigenvalues~\cite{criscione2000invariant}:  linear degenerate tensors correspond to $\mu = 1$, neutral tensors to $\mu = 0$, and planar degenerate tensors to $\mu = -1$.

A 3D tensor field is a continuous function $\mT(\vx)$ assigning a tensor to every location $\vx \in \RRSet^3$, while a tensor field ensemble is a set of $m$ such fields defined on the same domain.
Thus, $\mT_i(\vx)$ with $i \in [1,\dots,m]$ denotes the tensor at location $\vx$ in ensemble member $i$.
When assuming a Gaussian distribution of tensor values within the ensemble~\cite{basser2007spectral, zhang2017overview+, gerrits2019towards}, it can be summarized by its mean tensor $\bar{\mT} = \frac{1}{m} \sum_{i=1}^{m}\mT_i$ and the fourth-order covariance tensor $\Sigma$.
A similar approach for mode values enables statistical description using the mean mode $\bar{\mu} = \frac{1}{m} \sum_{i=1}^{m}\mu(\mT_i)$ and its standard deviation $\sigma$.

\section{Related Work}
Following the survey by Bi et al.~\cite{bi2019survey}, tensor field visualization can be broadly categorized into glyph-based approaches, which encode local tensor properties using geometric metaphors~\cite{schultz2010superquadric, gerrits2016glyphs}, and streamline-based approaches, including techniques such as hyperstreamlines~\cite{fu2014topologically}, tractography~\cite{CHAMBERLAND2025381}, and topological analysis.
Topological methods use tensor properties to reveal global structure and behavior~\cite{hesselink97topo} and remain an active area of research.
Notable contributions include the robust and efficient extraction of degenerate tensor locations~\cite{zheng2005topological}, neutral surfaces~\cite{palacios2015feature, roy2018robust}, and general mode surfaces~\cite{qu2021mode} alongside other topological descriptions~\cite{oster2018parallel, zobel2018extremal}.
Recent works have emphasized the relevance of mode surfaces and their transitions for comprehensive analysis~\cite{hung2023global, Lin2024asymp}, and have extended topology-based approaches to asymmetric tensor fields~\cite{hung2021feature}.\\
Visualizing uncertainty in data acquisition, simulation, or processing remains a major challenge in visualization~\cite{johnson2004top}.
For tensor data, uncertainty-aware approaches largely fall into the same categories: glyph-based techniques~\cite{zhang2017overview+, abbasloo2015visualizing, gerrits2019towards}, streamline-based methods like uncertain tractography~\cite{siddiqui2021progressive}, and the use of derived scalar descriptions to apply established uncertainty visualization techniques~\cite{pothkow2012uncertainty}.
In terms of topological features, there is a clear gap in uncertain visualization techniques.
Only recently, Schmitz and Gerrits~\cite{schmitz2024exploring} proposed features to capture uncertain degenerate tensor locations in symmetric tensor field ensembles.
Their method combines the tensor mode of both the mean tensor field and the distribution of mode values across the ensemble members.
The \textit{meanLine} is the degenerate tensor line extracted from the mean tensor field and acts as a representative of the ensemble.
The \textit{modeTube} builds upon the \textit{meanLine} but indicates the mode value distribution surrounding it, while the \textit{probabilityBand} is used to indicate the probability of mode value ranges within their local distribution.
While effective, their approach only focuses on degenerate tensor locations and does not distinguish tensor mode sign.
Further, they provide no guidance on how to extend their reasoning to general mode surfaces, leaving this feature unexplored.
In this work, we seek to address this gap.

\section{Uncertain Mode Surfaces in Tensor Ensembles}
Our goal is to generalize the uncertain degenerate tensor features introduced in \cite{schmitz2024exploring} to represent arbitrary mode surfaces.
Their features were designed with two goals in mind:
Effectively summarize the behavior of degenerate tensors within the full ensemble, and provide insights into the likelihood of the occurrence of degenerate tensors.
We aim to maintain these goals but replace the specific focus on degenerate tensors with arbitrary mode surfaces.
Degenerate tensor locations can be interpreted as $\pm1$-mode surfaces, i.e., special cases of mode surfaces that collapse into line structures.
Thus, we seek to extend the visualization techniques developed for line features to surfaces, while maintaining their ability to convey uncertainty.
In the following, we present the original definition of each feature, followed by its generalization.
All features rely on similar reasoning: combining mode information from both the mean tensor field and the distribution of mode values across the ensemble members.
However, unlike the original work, we explicitly consider the sign of the mode values and not just their absolute magnitude.

\subsection{Ensemble \textit{meanSurface}}
In prior work, the (enhanced) \textit{meanLine} was defined as the degenerate tensor line extracted from the ensemble mean tensor field.
It served as a representative of the whole ensemble and was augmented with additional quantities, such as mode standard deviation mapped to color or line thickness.
We generalize this idea from lines to surfaces by introducing the ensemble \textit{meanSurface} $\mathcal{M}_t$, which is the $t$-mode surface extracted from the mean tensor field:
\begin{equation}
    \mathcal{M}_t = \{\vx~|~\mu(\bar{\mT}(x)) = t\}.
\end{equation}
This extends the original formulation to arbitrary target mode values $t$.
Thus, if present in the data, this results in non-oriented surfaces for $t \neq \pm1$ and line structures for $t = \pm1$.
An \textit{enhanced meanSurface} is similarly achieved by augmenting the surface with additional information, e.g., using additional geometry or color to indicate local standard deviation of mode distribution.
While line thickness was used in the original work to indicate uncertainty, we generalize this by introducing a notion of ``surface thickness":
A line has an ambiguous normal direction in its orthogonal plane and uncertainty is visualized by offsetting the line radially by an amount determined by a scalar uncertainty value to form a tube.
In contrast, for surfaces, the normal vector $\vn(\vx)$ is unique up to sign.
Thus, the surface can be ``thickened" by offsetting each point in both forward and backward normal directions and connecting the resulting surfaces.
Note, however, that other techniques could be used to represent scalar values on surfaces.

\subsection{Ensemble \textit{modeShell}}
The \textit{meanSurface} encodes local properties but does not capture surrounding variation within the ensemble.
Schmitz and Gerrits addressed this for the ensemble \textit{meanLine} by introducing the ensemble \textit{modeTube}, sampling mode values around the \textit{meanLine} to create offset geometry reflecting spatial distribution of mode values. 
The tube expands or contracts depending on whether nearby mode values are closer or farther from the target mode, which for the original work was always $1$.
We generalize this concept to surfaces by sampling mode values at offset points in both surface normal directions for each point on the \textit{meanSurface} to determine whether the surrounding ensemble field trends toward or away from the target mode value $t$.
Let $\vp_0$ be a point on the \textit{meanSurface} with unit surface normal $\vn$.
We define offset points in positive and negative normal direction as $\vp_{c} = \vp_0 \pm \delta \cdot \vn$, where $\delta$ is a fixed step size.
At each offset point, we sample the mean mode value and evaluate how close it is to the target mode value using $d(\vp_c) = |(\bar{\mu}(\vp_0) - t)| - |(\bar{\mu}(\vp_c) - t)|$.
A positive $d(\vp_c)$ indicates that the mean mode at the offset point is closer to the target mode, while a negative value indicates that the value at the surface point is closer.
This comparison is visualized by shifting the surface points along their normals:
\begin{equation}
    \Tilde{\vp_c} = \vp_0 + \frac{2}{1 + e^{(-10 \, d(\vp_c))}} \cdot (\vp_{c} - \vp_0).
\end{equation}
An additional factor can be used to control minimum and maximum offset distances.
This offset representation, which originally only formed a tube around a line, now forms a shell around a surface, hence we refer to it as the \textit{modeShell}.
The user-defined offset distance $\delta$ replaces the radius from \cite{schmitz2024exploring}.
Distance and color indicate mode value distribution in the ensemble surrounding the \textit{meanSurface}.
Unlike the original \textit{modeTube}, our method allows arbitrary target mode values, making it broadly applicable beyond degenerate tensors.

\subsection{Generalized Ensemble \textit{probabilityBand}}
Both previous features rely on mode surfaces extracted from the mean tensor field, which might not always be present.
To overcome this limitation, Schmitz and Gerrits introduced the \textit{probabilityBand}, a feature surface derived directly from mode value distribution within all ensemble members.
Given a target probability $p$, it is defined as the isosurface $\mathcal{S}_p = \{\vx \in \RRSet^3 ~|~ f(\vx) = p\}$ where $f(\vx)$ describes the probability $P(X \geq b)_{\vx}$ for each location, that mode values exceed a certain threshold $t$ at $\vx$.
This original formulation prioritized mode values close to $1$, i.e., degenerate tensors.
To generalize this idea, we redefine the function $f(\vx)$ as the probability that mode values fall within a range around the target value $t$, which is typically identical to the mode value of interest:
\begin{equation}
    f(\vx) = P(t - \varepsilon_l \leq X \leq t + \varepsilon_u)
\end{equation}
where $\epsilon_u$ and $\epsilon_l$ define the acceptable deviation from the target mode value $t$.
Assuming a Gaussian distribution with mean $\bar{\mu}$ and standard deviation $\sigma$, this probability is computed via the cumulative distribution function $f(\vx) = \Phi(z_u)-\Phi(z_l)$ where
\begin{equation}
    z_l = \frac{(t - \epsilon_l) - \bar{\mu}}{\sigma}~\text{and}~z_u = \frac{(t + \epsilon_u) - \bar{\mu}}{\sigma}
\end{equation}
and approximated by the error function
\begin{equation}
    f(\vx) = \frac{1}{2} \left[ \operatorname{erf}\left( \frac{z_u}{\sqrt{2}} \right) - \operatorname{erf}\left( \frac{z_l}{\sqrt{2}} \right) \right].
\end{equation}
The \textit{generalized ensemble probabilityBand} is then the isosurface $\mathcal{S}_p$ of $f(\vx)$ for a user-defined probability $p$.
To recover the original degenerate case, one can set $\varepsilon_l = 0$ and $\varepsilon_u = 1 - t$ (or appropriately for $t = -1$).
While the generalization introduces additional parameters, it can now be used to indicate the appearance of arbitrary mode surfaces in the ensemble alongside their respective probability.

\section{Results}
We demonstrate the effectiveness of our novel techniques by extracting uncertain mode surface features in ensembles created from two real-world simulated datasets.
Reading, processing, and writing the data was performed using a custom C++ implementation based on the Visualization Toolkit (VTK)~\cite{vtkBook}.
Due to code availability, we used VTK to extract isocontours for the mode surfaces and \textit{probabilityBand} features, as this provided sufficiently accurate approximations of the features, and the open-source implementation by Oster et al.~\cite{oster2018parallel} for the extraction of degenerate tensor line locations.
Note that one could also implement the framework presented by Qu et al.~\cite{qu2021mode} to achieve more accurate feature geometry.

\subsection{Residual Stress in Additive Manufactured Materials}
\begin{figure}[ht!]
    \centering
    \includegraphics[width=0.84\linewidth]{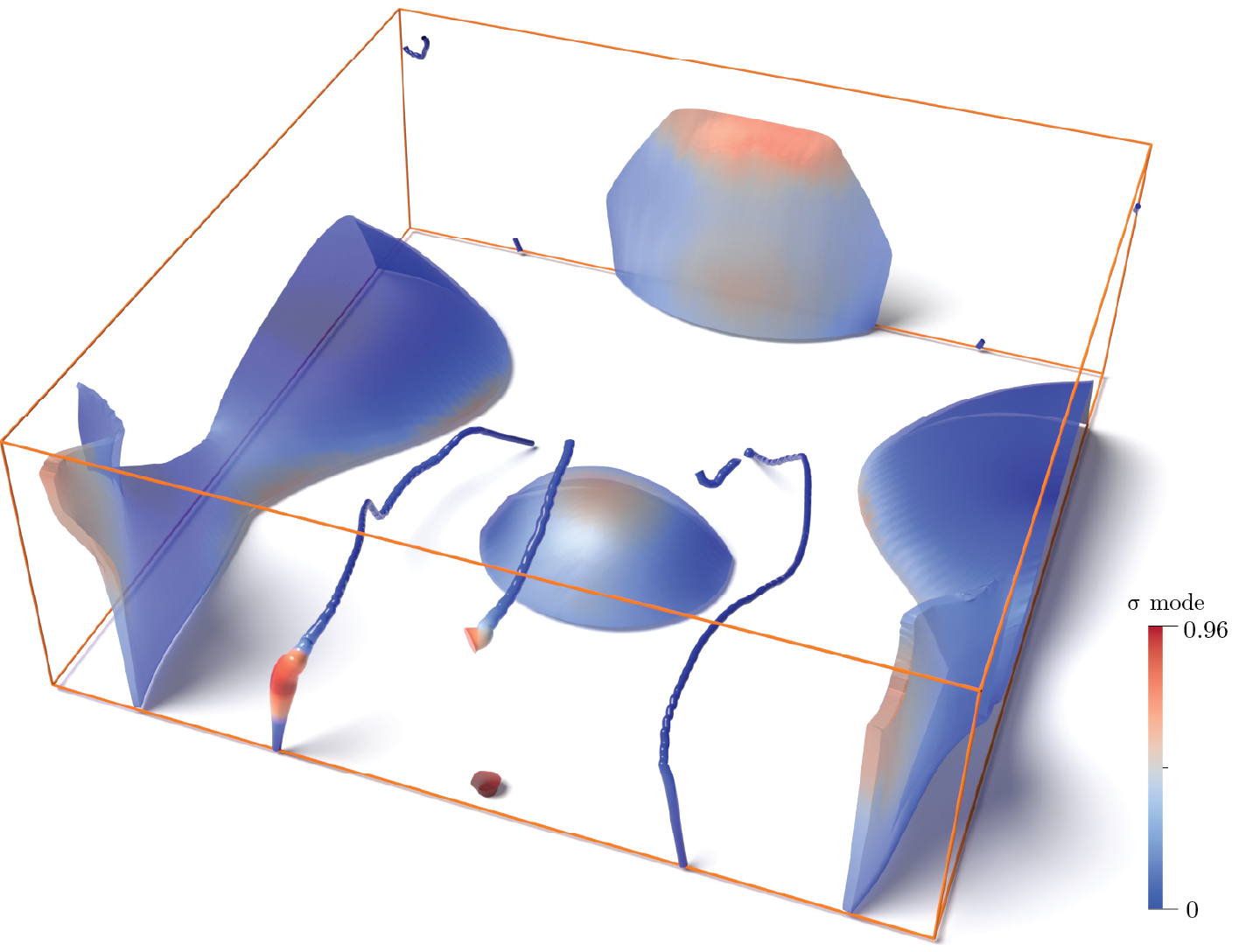}
    \caption{Combination of uncertain mode surface features: Both the uncertain neutral surface $\mathcal{M}_{0}$ and the uncertain planar degenerate tensor locations $\mathcal{M}_{-1}$ are shown, where color and thickness represent mode standard deviation.}
    \label{fig:mat_0_-1}
\end{figure}
The first dataset contains four ensemble members taken from a collection of residual stress simulations in additively manufactured materials~\cite{yang2024elasto}, publicly available~\cite{yang_2024_10940626}.
In a Selective Laser Sintering (SLS) process, successive layers of particles fuse and deform, introducing stress into the underlying substrate layer.
Each member represents adding a single layer to the simulation, as shown in \cref{fig:teaser}a), resulting in varying temperature and stress fields.
\cref{fig:teaser}b) shows the $-0.3$-mode surfaces extracted from each individual ensemble member, each depicted in a different shade of green.
In \cref{fig:teaser}c), the enhanced \textit{meanSurface} provides a compact summary that approximates the overall distribution of $-0.3$-mode surfaces while encoding uncertainty through mapping local standard deviation to shape and color.
The generalized ensemble \textit{probabilityBand} shown in \cref{fig:teaser}d) further highlights regions with a $33\%$ probability of mode values lying within the range of $-0.5$ to $-0.1$ offering a slightly different view of potential surface locations.
Similar to the work of Qu et al.~\cite{qu2021mode}, our approach supports the joint visualization of different uncertain mode surfaces.
\cref{fig:mat_0_-1} illustrates this by combining the uncertain neutral surface $\mathcal{M}_0$ with uncertain planar degenerate tensor locations $\mathcal{M}_{-1}$.
Both encode uncertainty using color and thickness and provide insights into the spatial variability and likely occurrence of mode structures across the ensemble.

\subsection{Simulation of Stresses in an O-Ring}
\begin{figure}[ht!]
    \centering
    \includegraphics[width=0.93\linewidth]{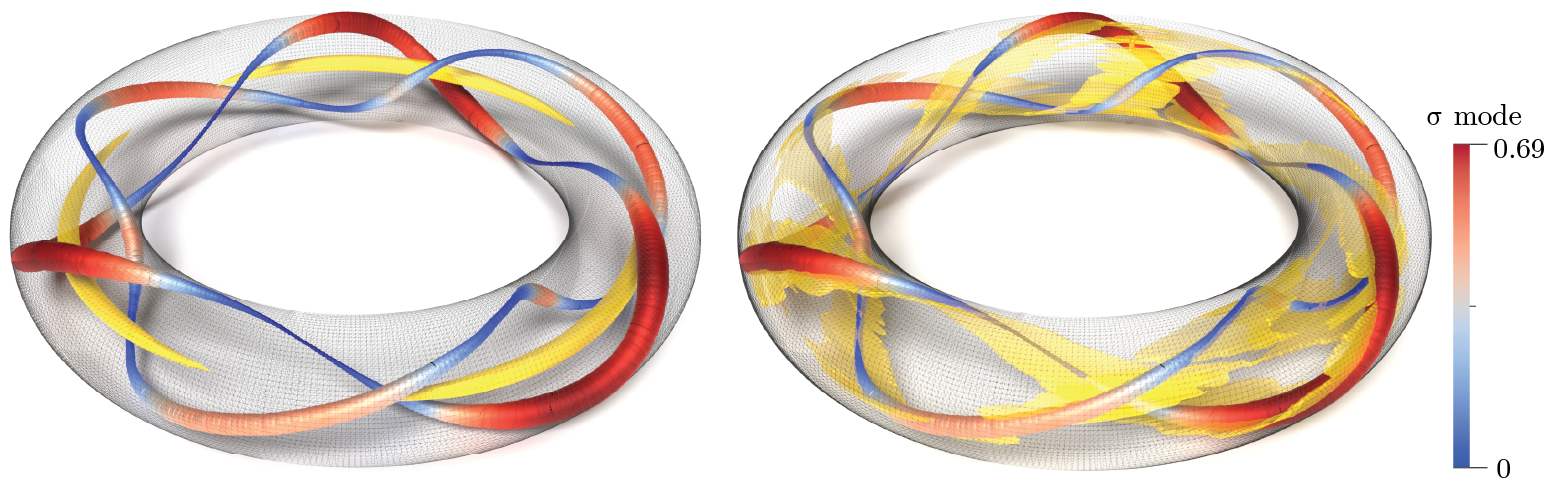}
    \caption{Degenerate tensor locations in a simulated O-ring stress ensemble under varying anisotropy $\alpha \in \{20, 25, 30, 33, 34\}$: generalized features (left) and original features~\cite{schmitz2024exploring} (right).  
    For $\mu = 1$, the generalized enhanced \textit{meanSurface} $\mathcal{M}_{1}$ visually resembles the enhanced \textit{meanLine} feature~\cite{schmitz2024exploring}, with thickness and color representing mode standard deviation.
    Unlike the \textit{meanLine}, however, the \textit{meanSurface} accounts for the sign of the mode.
    While the mean tensor field does not yield a surface for $\mu = -1$, the generalized \textit{probabilityBand} captures locations with a $36\%$ chance of mode values falling within $[-1.0, -0.8]$, which we can now distinguish.
    In contrast, the \textit{probabilityBand} (indicating a $36\%$ chance of mode values $\geq 0.8$) on the right fails to capture locations of values close to $-1$, as the distribution is dominated by the positive mode values.
    }
    \label{fig:ring_-1_1}
\end{figure}
\begin{figure*}[t!]
  \centering
    \begin{tikzpicture}
    \tikzstyle{image} = [inner sep=0, outer sep=0, node distance=0 and 0]
    \node[image] (image) at (0,0) {
        \includegraphics[width=0.97\linewidth]{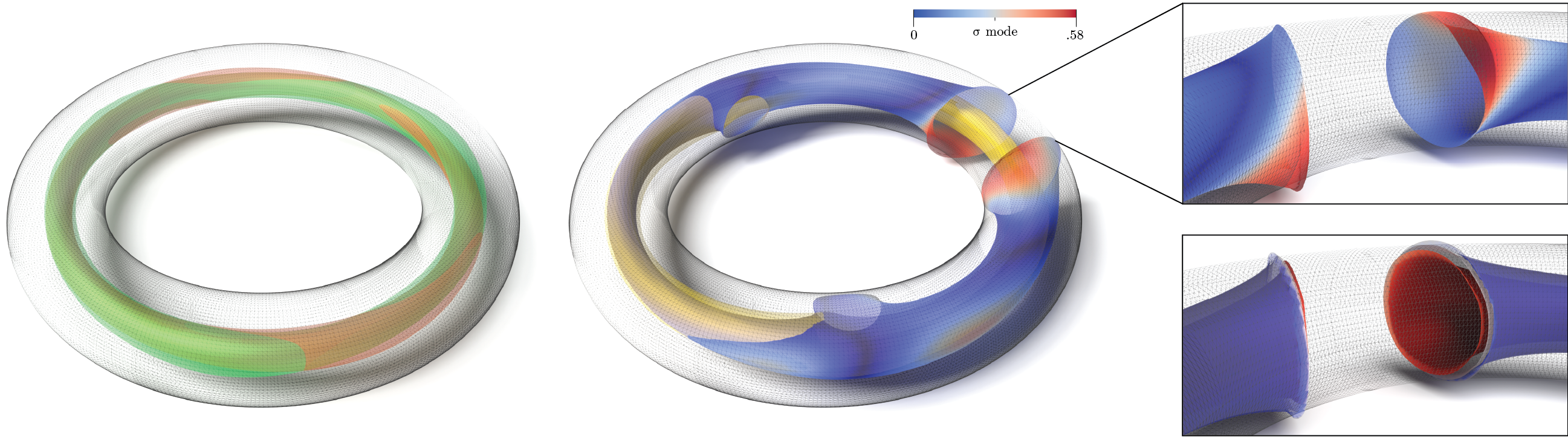}%
    };
    \begin{scope}[
        shift=(image.south west), 
	x={($(image.south east)-(image.south west)$)}, 
	y={($(image.north west)-(image.south west)$)}]
      \node[font=\small, text=black] at (0.17,-0.03) {a)};
      \node[font=\small, text=black] at (0.53,-0.03) {b)};
      \node[font=\small, text=black] at (0.88,0.5) {c)};
      \node[font=\small, text=black] at (0.88,-0.03) {d)};
    \end{scope}
  \end{tikzpicture}
  \caption{Uncertain $0.7$-mode surfaces in a two-member ensemble of simulated stresses in an O-ring with $p \in \{2,3\}$.
  a) $0.7$-mode surfaces extracted from each ensemble member.
  b) A composite visualization showing both the enhanced \textit{meanSurface} $\mathcal{M}_{0.7}$ where color and thickness indicate mode standard deviation, and the generalized \textit{probabilityBand} for a probability of $90\%$ and $\epsilon_l = \epsilon_u = 0.2$. 
  c) Closeup view of the enhanced \textit{meanSurface} $\mathcal{M}_{0.7}$ with thickness and color representing mode standard deviation.
  d) Closeup view of the ensemble \textit{modeShell} consisting of the \textit{meanSurface} $\mathcal{M}_{0.7}$ and two offset surfaces indicating the spatial distribution of mode values close to the \textit{meanSurface}.
  }
  \label{fig:RING}
\end{figure*}
The second dataset describes stresses in an O-ring under varying applied forces.
The simulation varies periodicity ($p,q$) and anisotropy magnitude ($\alpha$).
This data was not only used to showcase the original uncertain degenerate tensor lines features but also in other related works on tensor topology~\cite{hung2023global}, thus providing a convenient comparison.
We extract features on two different sets of simulated O-ring data, whereas the first comprises five members with varying anisotropy.
\cref{fig:ring_-1_1} (left) shows a notable difference from the original work (right), as it was previously not possible to distinguish between linear and planar degenerate tensor locations.
While the \textit{meanSurface} $\mathcal{M}_{1}$ results in a clear line structure that is similar to the uncertain \textit{meanLine}, the mean tensor field does not produce a \textit{meanSurface} for $\mu = -1$.
However, extracting the generalized \textit{probabilityBand} for a probability of $36\%$ of mode values in the range of $-1$ to $-0.8$ shown in yellow provides an indication of planar degenerate tensor locations in the ensemble.
The \textit{probabilityBand} of \cite{schmitz2024exploring} on the right is dominated by the positive mode values, which form around the \textit{meanLine}, and thus fail to provide insights into the location of tensors with $\mu = -1$.
The second ensemble contains two members where only periodicity is varied, such that both fields are very similar.
This is also clear from both $0.7$-mode surfaces extracted from the individual members in \cref{fig:RING}a) where only slight differences appear on the right half of the domain.
The enhanced \textit{meanSurface} $\mathcal{M}_{0.7}$ effectively captures the shared structure and highlights areas of variability via color and thickness, as shown in \cref{fig:RING} b) and c).
\cref{fig:RING}~d) visualizes the \textit{modeShell}, which provides additional insight into the spatial distribution of mode values.
The red offset surface shows that samples taken inside the shell are closer to the target value than those on the \textit{meanSurface} (gray), which is shown by the distance of the surfaces.
The blue offset surface indicates that sampled mode values just outside the shell are farther from the target, leading to both surfaces nestling to one another.  
Again, the composite visualization of both \textit{probabilityBand} and \textit{meanSurface} shown in \cref{fig:RING}b) can effectively summarize the mode surface distribution of the ensemble, whereas a single feature might fail to provide enough insight.

\section{Conclusion, Limitations, and Future Work}
In this work, we present a generalization of uncertain degenerate tensor features introduced by Schmitz and Gerrits~\cite{schmitz2024exploring} to arbitrary mode surfaces, introducing novel techniques for the visualization of uncertain features in 3D symmetric second-order tensor field ensembles.
Building on the original approach, our design incorporates both the information from the mean tensor field as an ensemble representative and the distribution of mode values across ensemble members.
A key enhancement is the explicit treatment of the mode sign, which was previously ignored.
When mode surfaces collapse to line structures, our method defaults to the original line-based features.
With these generalizations, we enable the analysis of behavior, transitions, and likelihoods of arbitrary mode surfaces within symmetric tensor field ensembles without additional computational costs.
We demonstrate our approach on real-world mechanical engineering and material science simulation ensembles.
While performance was not a focus in this work, the necessary processing of the data is achieved within seconds as it follows similar computations as the original work and can be easily run in parallel, providing a scalable technique for even very large ensembles.
Through this work, we aim to expand the set of tools available for the analysis of uncertain tensor field topology.

While the generalization to mode surfaces is a natural extension, several aspects deserve further discussion.
As the primary goal was a direct extension of the original topological concepts to surfaces, not all concepts are necessarily as suitable for surfaces as they have been for lines:
Surface-based features prove to be more challenging in terms of visual clutter, especially for convoluted surfaces or feature combinations, which becomes obvious for the \textit{modeShell} when applied to tube-shaped surfaces.
While using transparent rendering helps mitigate this, investigating more effective and scalable surface representations remains an open research area.
Further, generalizing the \textit{probabilityBand} introduces parameters controlling acceptable mode value ranges, which can add complexity to both interpretation and interaction.
Moreover, there already exist techniques for visualizing uncertain scalar fields, such as probabilistic marching cubes~\cite{athawale2021ms} or uncertainty-aware isocontour methods~\cite{pothkow2012uncertainty} that could potentially be applied to mode distributions.
Comparing such techniques to our direct generalization could lead to improved or alternative visual encodings.
Furthermore, our current framework retains parameters and assumptions from prior work, i.e., assuming Gaussian-distributed values, which may limit applicability in more complex scenarios and require similar parameter tuning as in the original work.
Extending our approach to asymmetric second-order tensor fields remains a significant challenge, especially given that symmetric tensor topology cannot be directly transferred~\cite{hung2021feature}.
Similarly, finding similar visual metaphors for higher-order tensor data is still an open and demanding research problem.
Finally, we plan to make the code publicly available. 

\acknowledgments{%
The authors gratefully acknowledge the German Federal Ministry of
Education and Research (BMBF) and the state government of NRW for
supporting this work/project as part of the NHR funding and thank Somnath Bharech for insights into domain challenges.
}

\bibliographystyle{abbrv-doi-hyperref}

\bibliography{template}

\begin{thebibliography}{10}

\bibitem{abbasloo2015visualizing}
A.~Abbasloo, V.~Wiens, M.~Hermann, and T.~Schultz.
\newblock Visualizing tensor normal distributions at multiple levels of detail.
\newblock {\em IEEE transactions on visualization and computer graphics}, 22(1):975--984, 2015.

\bibitem{athawale2021ms}
\href{https://doi.org/10.1109/VIS49827.2021.9623267}{T.~M. Athawale, S.~Sane, and C.~R. Johnson}.
\newblock \href{https://doi.org/10.1109/VIS49827.2021.9623267}{Uncertainty visualization of the marching squares and marching cubes topology cases}.
\newblock \href{https://doi.org/10.1109/VIS49827.2021.9623267}{In {\em 2021 IEEE Visualization Conference (VIS)}}, \href{https://doi.org/10.1109/VIS49827.2021.9623267}{pp. 106--110}, \href{https://doi.org/10.1109/VIS49827.2021.9623267}{2021}. \href{https://doi.org/10.1109/VIS49827.2021.9623267}
{doi: {{%
10\hspace{.1pt}\discretionary{.}{%
}{.}\hspace{.4pt}1109\discretionary{/}{%
}{/}VIS49827\hspace{.1pt}\discretionary{.}{%
}{.}\hspace{.4pt}2021\hspace{.1pt}\discretionary{.}{%
}{.}\hspace{.4pt}9623267}}}


\bibitem{basser2007spectral}
P.~J. Basser and S.~Pajevic.
\newblock Spectral decomposition of a 4th-order covariance tensor: Applications to diffusion tensor mri.
\newblock {\em Signal Processing}, 87(2):220--236, 2007.

\bibitem{bi2019survey}
C.~Bi, L.~Yang, Y.~Duan, and Y.~Shi.
\newblock A survey on visualization of tensor field.
\newblock {\em Journal of Visualization}, 22:641--660, 2019.

\bibitem{CHAMBERLAND2025381}
\href{https://doi.org/https://doi.org/10.1016/B978-0-12-818894-1.00002-1}{M.~Chamberland, C.~Poirier, T.~Hendriks, D.~Shastin, A.~Vilanova, and A.~Leemans}.
\newblock \href{https://doi.org/https://doi.org/10.1016/B978-0-12-818894-1.00002-1}{Chapter 20 - tractography visualization}.
\newblock \href{https://doi.org/https://doi.org/10.1016/B978-0-12-818894-1.00002-1}{In F.~Dell'Acqua, M.~Descoteaux, and A.~Leemans, eds., {\em Handbook of Diffusion MR Tractography}}, \href{https://doi.org/https://doi.org/10.1016/B978-0-12-818894-1.00002-1}{pp. 381--393}. \href{https://doi.org/https://doi.org/10.1016/B978-0-12-818894-1.00002-1}{Academic Press}, \href{https://doi.org/https://doi.org/10.1016/B978-0-12-818894-1.00002-1}{2025}. \href{https://doi.org/10.1016/B978-0-12-818894-1.00002-1}
{doi: {{%
10\hspace{.1pt}\discretionary{.}{%
}{.}\hspace{.4pt}1016\discretionary{/}{%
}{/}B978\discretionary{%
}{-}{-}0\discretionary{%
}{-}{-}12\discretionary{%
}{-}{-}818894\discretionary{%
}{-}{-}1\hspace{.1pt}\discretionary{.}{%
}{.}\hspace{.4pt}00002\discretionary{%
}{-}{-}1}}}


\bibitem{criscione2000invariant}
J.~C. Criscione, J.~D. Humphrey, A.~S. Douglas, and W.~C. Hunter.
\newblock An invariant basis for natural strain which yields orthogonal stress response terms in isotropic hyperelasticity.
\newblock {\em Journal of the Mechanics and Physics of Solids}, 48(12):2445--2465, 2000.

\bibitem{fu2014topologically}
F.~Fu and N.~M. Abukhdeir.
\newblock A topologically-informed hyperstreamline seeding method for alignment tensor fields.
\newblock {\em IEEE Transactions on Visualization and Computer Graphics}, 21(3):413--419, 2014.

\bibitem{gerrits2016glyphs}
T.~Gerrits, C.~R{\"o}ssl, and H.~Theisel.
\newblock Glyphs for general second-order 2d and 3d tensors.
\newblock {\em IEEE transactions on visualization and computer graphics}, 23(1):980--989, 2016.

\bibitem{gerrits2019towards}
T.~Gerrits, C.~R{\"o}ssl, and H.~Theisel.
\newblock Towards glyphs for uncertain symmetric second-order tensors.
\newblock In {\em Computer Graphics Forum}, vol.~38, pp. 325--336. Wiley Online Library, 2019.

\bibitem{hesselink97topo}
\href{https://doi.org/10.1109/2945.582332}{L.~Hesselink, Y.~Levy, and Y.~Lavin}.
\newblock \href{https://doi.org/10.1109/2945.582332}{The topology of symmetric, second-order 3d tensor fields}.
\newblock \href{https://doi.org/10.1109/2945.582332}{{\em IEEE Transactions on Visualization and Computer Graphics}}, \href{https://doi.org/10.1109/2945.582332}{3(1):1--11}, \href{https://doi.org/10.1109/2945.582332}{1997}. \href{https://doi.org/10.1109/2945.582332}
{doi: {{%
10\hspace{.1pt}\discretionary{.}{%
}{.}\hspace{.4pt}1109\discretionary{/}{%
}{/}2945\hspace{.1pt}\discretionary{.}{%
}{.}\hspace{.4pt}582332}}}


\bibitem{hung2021feature}
S.-H. Hung, Y.~Zhang, H.~Yeh, and E.~Zhang.
\newblock Feature curves and surfaces of 3d asymmetric tensor fields.
\newblock {\em IEEE Transactions on Visualization and Computer Graphics}, 28(1):33--42, 2021.

\bibitem{hung2023global}
S.-H. Hung, Y.~Zhang, and E.~Zhang.
\newblock Global topology of 3d symmetric tensor fields.
\newblock {\em IEEE Transactions on Visualization and Computer Graphics}, 30(1):1282--1291, 2023.

\bibitem{johnson2004top}
C.~Johnson.
\newblock Top scientific visualization research problems.
\newblock {\em IEEE Computer Graphics and Applications}, 24(4):13--17, 2004.

\bibitem{kamal2021recent}
A.~Kamal, P.~Dhakal, A.~Y. Javaid, V.~K. Devabhaktuni, D.~Kaur, J.~Zaientz, and R.~Marinier.
\newblock Recent advances and challenges in uncertainty visualization: a survey.
\newblock {\em Journal of Visualization}, 24(5):861--890, 2021.

\bibitem{Lin2024asymp}
\href{https://doi.org/10.1109/TopoInVis64104.2024.00010}{X.~Lin, Y.~Zhang, and E.~Zhang}.
\newblock \href{https://doi.org/10.1109/TopoInVis64104.2024.00010}{Asymptotic topology of 3d linear symmetric tensor fields}.
\newblock \href{https://doi.org/10.1109/TopoInVis64104.2024.00010}{In {\em 2024 IEEE Topological Data Analysis and Visualization (TopoInVis)}}, \href{https://doi.org/10.1109/TopoInVis64104.2024.00010}{pp. 55--64}, \href{https://doi.org/10.1109/TopoInVis64104.2024.00010}{2024}. \href{https://doi.org/10.1109/TopoInVis64104.2024.00010}
{doi: {{%
10\hspace{.1pt}\discretionary{.}{%
}{.}\hspace{.4pt}1109\discretionary{/}{%
}{/}TopoInVis64104\hspace{.1pt}\discretionary{.}{%
}{.}\hspace{.4pt}2024\hspace{.1pt}\discretionary{.}{%
}{.}\hspace{.4pt}00010}}}


\bibitem{oster2018parallel}
T.~Oster, C.~R{\"o}ssl, and H.~Theisel.
\newblock The parallel eigenvectors operator.
\newblock In {\em VMV}, pp. 39--46, 2018.

\bibitem{palacios2015feature}
J.~Palacios, H.~Yeh, W.~Wang, Y.~Zhang, R.~S. Laramee, R.~Sharma, T.~Schultz, and E.~Zhang.
\newblock Feature surfaces in symmetric tensor fields based on eigenvalue manifold.
\newblock {\em IEEE transactions on visualization and computer graphics}, 22(3):1248--1260, 2015.

\bibitem{pothkow2012uncertainty}
K.~P{\"o}thkow and H.-C. Hege.
\newblock Uncertainty propagation in dt-mri anisotropy isosurface extraction.
\newblock In {\em New Developments in the Visualization and Processing of Tensor Fields}, pp. 209--225. Springer, 2012.

\bibitem{qu2021mode}
\href{https://doi.org/10.1109/TVCG.2020.3030431}{B.~Qu, L.~Roy, Y.~Zhang, and E.~Zhang}.
\newblock \href{https://doi.org/10.1109/TVCG.2020.3030431}{Mode surfaces of symmetric tensor fields: Topological analysis and seamless extraction}.
\newblock \href{https://doi.org/10.1109/TVCG.2020.3030431}{{\em IEEE Transactions on Visualization and Computer Graphics}}, \href{https://doi.org/10.1109/TVCG.2020.3030431}{27(2):583--592}, \href{https://doi.org/10.1109/TVCG.2020.3030431}{2021}. \href{https://doi.org/10.1109/TVCG.2020.3030431}
{doi: {{%
10\hspace{.1pt}\discretionary{.}{%
}{.}\hspace{.4pt}1109\discretionary{/}{%
}{/}TVCG\hspace{.1pt}\discretionary{.}{%
}{.}\hspace{.4pt}2020\hspace{.1pt}\discretionary{.}{%
}{.}\hspace{.4pt}3030431}}}


\bibitem{roy2018robust}
L.~Roy, P.~Kumar, Y.~Zhang, and E.~Zhang.
\newblock Robust and fast extraction of 3d symmetric tensor field topology.
\newblock {\em IEEE transactions on visualization and computer graphics}, 25(1):1102--1111, 2018.

\bibitem{schmitz2024exploring}
T.~Schmitz and T.~Gerrits.
\newblock Exploring uncertainty visualization for degenerate tensors in 3d symmetric second-order tensor field ensembles.
\newblock In {\em 2024 IEEE Workshop on Uncertainty Visualization: Applications, Techniques, Software, and Decision Frameworks}, pp. 1--11. IEEE Computer Society, 2024.

\bibitem{vtkBook}
W.~Schroeder, K.~Martin, and B.~Lorensen.
\newblock {\em The Visualization Toolkit (4th ed.)}.
\newblock Kitware, 2006.

\bibitem{schultz2010superquadric}
T.~Schultz and G.~L. Kindlmann.
\newblock Superquadric glyphs for symmetric second-order tensors.
\newblock {\em IEEE transactions on visualization and computer graphics}, 16(6):1595--1604, 2010.

\bibitem{siddiqui2021progressive}
F.~Siddiqui, T.~H{\"o}llt, and A.~Vilanova.
\newblock A progressive approach for uncertainty visualization in diffusion tensor imaging.
\newblock In {\em Computer graphics forum}, vol.~40, pp. 411--422. Wiley Online Library, 2021.

\bibitem{yang_2024_10940626}
\href{https://doi.org/10.5281/zenodo.10940626}{Y.~Yang and S.~Bharech}.
\newblock \href{https://doi.org/10.5281/zenodo.10940626}{Elasto-plastic residual stress analysis of selective laser sintered porous materials based on 3d-multilayer thermo-structural phase-field simulations}, \href{https://doi.org/10.5281/zenodo.10940626}{Apr. 2024}. \href{https://doi.org/10.5281/zenodo.10940626}
{doi: {{%
10\hspace{.1pt}\discretionary{.}{%
}{.}\hspace{.4pt}5281\discretionary{/}{%
}{/}zenodo\hspace{.1pt}\discretionary{.}{%
}{.}\hspace{.4pt}10940626}}}


\bibitem{yang2024elasto}
Y.~Yang, S.~Bharech, N.~Finger, X.~Zhou, J.~Schr{\"o}der, and B.-X. Xu.
\newblock Elasto-plastic residual stress analysis of selective laser sintered porous materials based on 3d-multilayer thermo-structural phase-field simulations.
\newblock {\em npj Computational Materials}, 10(1):117, 2024.

\bibitem{zhang2017overview+}
C.~Zhang, M.~W. Caan, T.~H{\"o}llt, E.~Eisemann, and A.~Vilanova.
\newblock Overview+ detail visualization for ensembles of diffusion tensors.
\newblock In {\em Computer Graphics Forum}, vol.~36, pp. 121--132. Wiley Online Library, 2017.

\bibitem{zheng2005topological}
X.~Zheng, B.~Parlett, and A.~Pang.
\newblock Topological structures of 3d tensor fields.
\newblock In {\em VIS 05. IEEE Visualization, 2005.}, pp. 551--558. IEEE, 2005.

\bibitem{zobel2018extremal}
V.~Zobel and G.~Scheuermann.
\newblock Extremal curves and surfaces in symmetric tensor fields.
\newblock {\em The Visual Computer}, 34(10):1427--1442, 2018.

\end{thebibliography}


\end{document}